\documentclass[runningheads]{llncs}
\usepackage{graphicx}
\usepackage{changepage}
\usepackage{lipsum}
\usepackage{xcolor}
\usepackage[breaklinks, colorlinks, allcolors=blue]{hyperref}
\usepackage{floatrow}
\usepackage{tabularx}
\usepackage{amssymb}
\usepackage{amsmath}
\usepackage{mathrsfs}
\usepackage[normalem]{ulem}  
\usepackage{hyperref}
\usepackage{url}
\usepackage{ctable} 
\usepackage{tabularx}
\usepackage{adjustbox}
\usepackage{booktabs}
\usepackage{array}
\usepackage{listings}
\usepackage{verbatim}

\usepackage{refcount}

\usepackage[nocompress]{cite} 

\usepackage[ampersand]{easylist}
\newcolumntype{?}{!{\vrule width 1pt}}
\newcommand{\rt}[1]{{\textcolor{black}{{#1}}}}

\newcommand{\arxiv}{ar$\rm{\chi}$iv }

\makeatletter
\def\thickhline{%
  \noalign{\ifnum0=`}\fi\hrule \@height \thickarrayrulewidth \futurelet
  \reserved@a\@xthickhline}
\def\@xthickhline{\ifx\reserved@a\thickhline
              \vskip\doublerulesep
              \vskip-\thickarrayrulewidth
             \fi
      \ifnum0=`{\fi}}
\makeatother

\newlength{\thickarrayrulewidth}
\usepackage{bigfoot}
\DeclareNewFootnote[para]{default}

\begin{document}


\title{Large Synthetic Data from the \arxiv for OCR Post Correction of Historic Scientific Articles \vspace{-3mm}}

\titlerunning{OCR post correction data from the \arxiv}
%
\author{J. P. Naiman\inst{1}\orcidID{0000-0002-9397-6189} \and 
Morgan G. Cosillo\inst{1}\orcidID{0009-0008-4755-7531} \and
Peter K. G. Williams\inst{2}\orcidID{0000-0003-3734-3587} \and
Alyssa Goodman\inst{2}\orcidID{0000-0003-1312-0477}}
\authorrunning{J.P. Naiman et al.}
%
\institute{School of Information Sciences, University of Illinois, Urbana-Champaign, 61820, USA
\email{\{jnaiman,mcosil2\}@illinois.edu}\\
\and
Harvard-Smithsonian Center for Astrophysics, Cambridge, 02138, USA\\
\email{\{pwilliams,agoodman\}@cfa.harvard.edu}}
\maketitle              
\vspace*{-8mm}
\begin{abstract}
    Historical scientific articles often require Optical Character Recognition (OCR) to transform scanned documents into machine-readable text, a process that often produces errors.  
    We present a pipeline for the generation of a synthetic ground truth/OCR dataset to correct the OCR results of the astrophysics literature holdings of the NASA Astrophysics Data System (ADS).  By mining the \arxiv we create, to the authors' knowledge, the largest scientific synthetic ground truth/OCR post correction dataset of \input{tolatex/total_aligned_characters.dat}\unskip character pairs.  Baseline models trained with this dataset find the mean improvement in character and word error rates of \rt{\input{tolatex/byt5_full_large_fix_historicalmean_cer_imp.dat}\unskip\% and \input{tolatex/byt5_full_large_fix_historicalmean_wer_imp.dat}\unskip\%} for historical OCR text, respectively.  
    Interactive dashboards to explore the dataset are available online: \url{https://readingtimemachine.github.io/projects/1-ocr-groundtruth-may2023}, and data and code, are hosted on GitHub: \url{https://github.com/ReadingTimeMachine/ocr_post_correction}.

\end{abstract}

\vspace*{-11mm}
\section{Introduction}
\vspace*{-2mm}
  The ability to digitally store and parse scientific literature is vital to ensure access and proliferation of scientific ideas \cite{mayernik2017,sandy2017,sohmen2018figures}.
  While digital storage is supported for much of contemporary scientific literature, the text of many historical documents is ``trapped" within scanned pages of paper journals and theses.  
  
  Recently, various deep learning methods have been employed to extract page objects (e.g., figures) from scans \cite{adsass2012,adsass2015,scanbankthesis,ahuja-etal-2022-parsing}.
  An obstacle to the extraction of information from historical articles is the accuracy of these extracted materials.
  This is especially of concern for any text objects which contain the bulk of the information in an article.
  A typical solution is to extract text with Optical Character Recognition (OCR) engines \cite{tafti2016ocr}.
  However, the generated text is often noisy which is not only an issue for comprehension by humans and screen readers \cite{schmitt-koopmann_accessible_2022}, but also can affect ``downstream" natural language processing tasks such as topic modeling, sentence segmentation and named entity recognition \cite{boros_assessing_2022}, often times causing significant errors in these processes \cite{van2020assessing}.

  Here, we discuss a new method for addressing OCR noise in the context of the extraction of text from a subset of $\sim$56k articles from the pre-digital holdings of the the Astrophysics Data System (ADS)\footnote{\url{https://ui.adsabs.harvard.edu/}} from $\sim$1850-1997 \cite{naiman2023digitization}.  
  While our ultimate goal is to correct all historical text within the ADS holdings, our initial focus is on the correction of ``plain text" in the main portions of articles (i.e., not text within tables or captions).
  Our method relies on generating synthetic data from mining the \arxiv source files (LaTeX/TeX files which compile to PDFs \cite{urban1986introduction}) for ``post correction" models which are applied to previously extracted OCR text.
  
  Post correction methods are vital to the extraction of text from the historical holdings of ADS as only a small portion of the articles can be mined with PDF-parsing software \cite{naiman2022figure,naiman2023digitization}.  Additionally, in many large historical corpora it is not computationally feasible to re-OCR holdings each time an OCR engine is upgraded \cite{zaytsev2015hathitrust}, making post correction the only option to reduce errors.
  
  While the work presented here focuses on the literature of the ``big-data" science of astronomy and astrophysics \cite{astrobigdata1,astrobigdata2}, our methods of synthetic data generation can be generalized to other scientific fields.
  To aid in future generalizability, we use the open-source OCR engine \textsf{Tesseract} \cite{tesseract} and provide all code in \textsf{Python}. 
  Because the dataset is large we provide interactive visualizations to assist any user of our resource in their investigation of the dataset.


\vspace*{-4mm}
\section{OCR Noise Reduction Techniques \& Mining the \arxiv}
\vspace*{-3mm}
OCR noise is prevalent in the majority of OCR datasets used in the fields of digital humanities and cultural analytics \cite{jiang2021gutenberg}.
OCR errors do not follow patterns of typical misspellings, thus their correction generally relies on different tools than spell-checking software \cite{nguyen_deep_2019}.
OCR post correction, a method of error mitigation, in which OCR'd text is de-noised, is a field covering a wide range of digitization applications \cite{rigaud_icdar_nodate} and  models have historically taken several forms \cite{zhu_novel_2016}.
More recently, deep learning models have been developed to tackle post correction \cite{maheshwari_benchmark_2022} which typically make use of sequence-to-sequence models \cite{xue_byt5_2022,ramirez2022post,liu2020mBART}.

These deep learning methods require large training datasets, making their testing predominately completed with well known OCR post correction datasets from the community \cite{icdar2017postcorr,icdar2019postcorr,overproofdata}.
As manual annotations can be time consuming at scale \cite{maheshwari_benchmark_2022,springmann2018ground}, synthetic datasets are often used \cite{li_tablebank_2020,maskrcnnDocbank,zharikov_ddi-100-2020}.
In particular, mining the \arxiv is a popular method to generate synthetic machine learning training datasets \cite{li_tablebank_2020,pfahler_self-supervised_2022,maskrcnnDocbank}.  Given the variety of journals represented in the \arxiv database, its mining represents a vital opportunity to create domain-specific synthetic data  \cite{rensynthOCRdata2016,krishnan2016generating,le2017using}, which is necessary as models trained on one type of document will often fail on documents dissimilar to the training data \cite{etter_synthetic_2019}.

\vspace*{-5mm}
\section{Methods}
\vspace*{-3mm}

In what follows, we make use of two decades of the oldest articles available through the \arxiv Bulk Downloads \cite{arxivbulk} (1991-2011) for a total of \input{tolatex/total_articles_in_timerange_arxiv.dat}\unskip articles.

\vspace{-3mm} 
\subsection{Compiling the Astrophysics \arxiv Bulk Downloads} 
\label{section:astroArts}
\vspace{-2mm}

Once downloaded, all article files are checked for corrupt decompression formats and a main TeX file (those containing \textsf{\textbackslash documentclass} or \textsf{\textbackslash documentstyle}) for a total of \input{tolatex/total_astro_files_with_class_or_style.dat}\unskip articles.
To construct an ``astronomy article" list, class/style commands are parsed with \textsf{regex} and those which denote typical astrophysical journal names (e.g., ``aastex", ``apj", ``mn") are kept. These names correspond to the three journals which have the most complete scanned historical corpus (The Astrophysical Journal, Astronomy \& Astrophysics, and Monthly Notices of the Royal Astronomical Society) \cite{eichhorn_new_2002}.
This results in a total of \input{tolatex/total_astronomy_articles_to_test_for_compilation.dat}\unskip articles. 


This set of $\sim$65k files are tested for PDF-compilation errors for a total of \input{tolatex/total_astronomy_articles_compiled_successfully.dat}\unskip successfully compiled astronomy articles.  
The main sources of error are missing files (e.g., missing figure files) and an inability to distinguish which TeX file in a directory is the main article document.

\vspace*{-4mm} 
\subsection{Segmentation of TeX Documents} \label{section:segmenting}
\vspace*{-2mm} 

Many parsers exist for TeX files with output formats such as plain text (e.g., \textsf{opendetex} \cite{opendetex}), XML (e.g., \textsf{LaTeXML} \cite{ginev_latexml_2014}, \textsf{unarXive} \cite{unarxiv2022}) or document trees (e.g., \textsf{TexSoup} \cite{texsoup}).
With all methods, this parsing tends to be non-trivial \cite{saier2019bibliometric}.
As the documents are compiled once marking modifications are applied to the TeX to track synthetic ground truth (SGT) locations, any parser must account for errors that could occur in the compilation process.  
Additionally, checks for incorrect splitting of TeX source into trees are required.
%
This excludes ``off the shelf" parsers which only run a subset of these checks\footnote{For example, following the process in \autoref{section:marking}, TeXSoup finds errors in only \input{tolatex/percent_error_fortikz_TeXSoup.dat}\unskip\% of files, while our method finds errors in \input{tolatex/percent_error_tikzmarked_successfully.dat}\unskip\%.}. Thus this work makes use of a custom-built TeX parser.


\autoref{fig:treeflow} diagrams the segmentation process which uses \textsf{regex} to break TeX files into document trees. 
A raw TeX document (``Raw LaTeX" snippet shown in upper left gray panel) is parsed to find the locations of special characters denoting commands, variables, and environments (``Splits with regex" blue upper middle panel).  
A hierarchy is then constructed with checks for closing and opening statements of commands (closing \{\}) inline math formulas (paired \$'s) and environments (\textbackslash begin, \textbackslash end) and stored in a tree (``Tree" purple upper right panel).
Commands which reside within plain text sentences such as inline math, citations, and references (``\textbackslash ref\{\}" commands) are stored with special tags.

\vspace{-3mm}
\subsection{Marking the ``ground truth" words in LaTeX \& OCR'ing Pages} \label{section:marking}
\vspace*{-2mm}
Many methods for marking TeX documents to generate synthetic data for page objects (e.g., figures) modify the LaTeX to add bounding boxes in specific colors around objects and use image processing techniques to extract object locations after the PDF is rendered \cite{li_tablebank_2020,maskrcnnDocbank}. 
Rendered PDFs can potentially be mined for SGT text, however, this can lead to errors in the extracted SGT text \cite{maskrcnnDocbank}.

\begin{figure}
\centering
\includegraphics[width=1\textwidth]{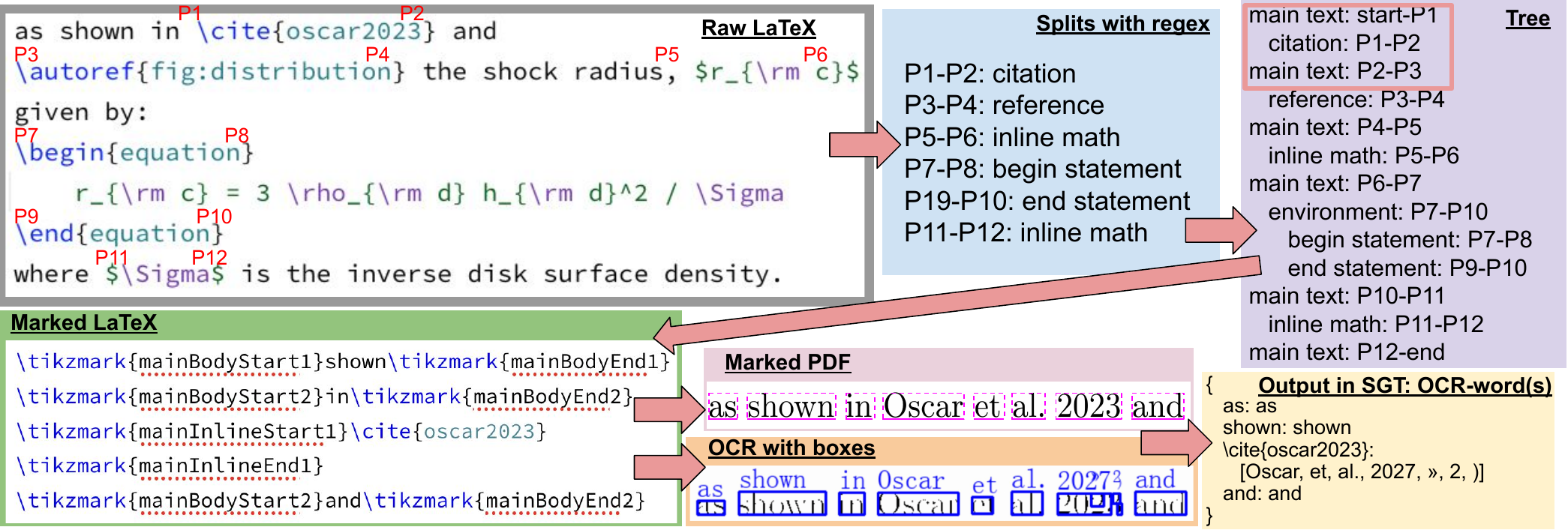}
\caption{Diagram of TeX parsed into its attributes (``Raw LaTeX", ``Splits with regex"), and the tree structure built from the positions of these splits within the document (``Tree"), as outlined in \autoref{section:segmenting}. TeX is then marked with the \textsf{tikzmark} package and OCR'd (section from three top lines in ``Tree" shown in ``Marked LaTeX", \autoref{section:marking}).  Once the TeX is compiled into a PDF, the auxiliary files are parsed to locate the SGT word locations on the rendered PDF page (``Marked PDF", \autoref{section:alignment}), OCR words are collected (``OCR with boxes"), and SGT-OCR boxes are aligned (``Output data SGT: OCR-word(s)", \autoref{section:alignment}). See text for more details.}
\label{fig:treeflow}
\end{figure}

To avoid SGT-text parsing errors, this work adopts a different approach by modifying the TeX source documents with markers denoting every word, inline equation, citation, and reference using the \textsf{tikzmark} \cite{tikzmark} package as shown within the green outlined ``Marked LaTeX" box of \autoref{fig:treeflow}.
Inline math, citations, and references are included as they are frequently interspersed with the plain text.

After storing the locations of each SGT object (``Tree" purple box in \autoref{fig:treeflow}), all text within the ``plain text" sections are split into words using white space and starting (ending) \textsf{\textbackslash tikzmark} commands are placed at the word/citation/reference/inline math start (end).  
Once the TeX document is compiled, the marks are stored in the auxilary (.aux) file produced during compilation which is then parsed to match each word to its location on the final, rendered PDF page.
At this stage, documents which contain the \textsf{\textbackslash input} command are ignored as these can include text external to the document being parsed.

Once the marked files are compiled, each page of each article is OCR'd with \textsf{Tesseract}, following methods used with articles from the historical holdings of the ADS \cite{naiman2022figure,naiman2023digitization}.  Examples of these bounding boxes and words are shown in the orange ``OCR with boxes" panel of \autoref{fig:treeflow}.
\vspace{-5mm}

\vspace{-12mm}
\subsection{The OCR-SGT Alignment Algorithm \& Dataset Characteristics} \label{section:alignment}
\vspace*{-2mm}
The final step in creating our SGT - OCR dataset is to align the OCR and SGT words.  In what follows, ``element" is defined as a plain text word, inline math formula, citation, or reference.  Our alignment routine is as follows:
\vspace*{-2mm}
\begin{itemize}
    \item Step 1: Locations of the bottom left and right bounds of each marked element are found from the .aux files.  These locations are shown as solid magenta lines in the magenta ``Marked PDF" panel of \autoref{fig:treeflow}.
    \item Step 2: As \textsf{tikzmark} gives only the lower y-position of each element, a bounding box is created by assuming 11pt font for each element (11pt font is an average value, font size is not always specified explicitly in the TeX file), shown by the dashed magenta lines in the ``Marked PDF" panel of \autoref{fig:treeflow}.
    \item Step 3: If the bounding box is found to span more than one line, the SGT element is assumed to be hyphenated and each part is marked as a separate word. Alignment operates page-by-page, therefore hyphenated elements which span multiple pages are ignored.
    \item Step 4: The ``raw" SGT element is extracted from the source TeX.
    \item Step 5: All OCR bounding boxes which overlap with a SGT box are associated with that SGT element.  If an OCR bounding box is associated with more than one SGT element, 
    the OCR element is associated with the SGT element with which it has the largest intersection-over-union (IOU).
    \item Step 6: All OCR elements associated with a SGT element are ordered by increasing horizontal position and combined into a single OCR element for that SGT element.  This is shown by the data structure in the yellow ``Output in SGT: OCR-words" box of \autoref{fig:treeflow}.
    \item Step 7: SGT word ``type" is stored along with SGT word (plain text, inline math, citation, reference and whether the word is hyphenated).
    \item Step 8: Elements are ordered by \textsf{tikzmark}s and aligned with edit distance operations \cite{Levenshtein2}. \textsf{spaCY} \cite{spacy2} is used to tokenize aligned pages as sentences \cite{spacysentence}.

\end{itemize}

While the majority of articles are aligned without error, \textsf{Tesseract} errors are possible on single pages.  
From this corpus of \input{tolatex/total_tikzmarked_files_without_not_all_errors.dat}\unskip articles which contain successfully aligned pages, our algorithm produces a total of 
    \input{tolatex/total_aligned_pages.dat}\unskip pages of \input{tolatex/total_aligned_sentences.dat}\unskip SGT/OCR sentence pairs which contain a total of \input{tolatex/total_aligned_characters.dat}\unskip character pairs. 

The relationships between SGT and OCR aligned characters closely follow other popular datasets with the majority of Levenshtein edit distance \cite{levenshtein} operations in our dataset (other datasets) being replacements $\sim$61.5\% ($\sim$40-60\%), followed by deletions $\sim$19.6\% ($\sim$10-18\%) and insertions $\sim$18.9\% ($\sim$5-24\%) \cite{nguyen_deep_2019}.
Interactive versions of large confusion matrices for alphabetic characters, digits, punctuation marks and frequent words are hosted on this project's webpage\footnote{\label{interactive}\url{https://readingtimemachine.github.io/projects/1-ocr-groundtruth-may2023}}.

\vspace{-4mm}
\section{Post Correction Model Baseline Tests}
\vspace*{-4mm}

To test the post correction effectiveness of our dataset we train a baseline transformer model --  \textsf{byt5} \cite{xue_byt5_2022} -- with the dataset. 
This model is effective for datasets such as ours which contain many out-of-vocabulary OCR words \cite{maheshwari_benchmark_2022}.
The model's initial training uses 100k aligned sentences for training, and 5k in the validation and test datasets.
Here, transfer learning from the \rt{byt5/google-small} model on HuggingFace \cite{huggingfacebyt5} is used, and, for all models, training occurs on a NVIDIA V100 for $\sim$87000 iterations over $\sim$24 hours, in which the model converges.

The entry above the first thick line of \autoref{tab:baseline} (``\textsf{byt5},words") shows the ability of the model to correct only the parts of each aligned SGT-OCR text which have been tagged as plain text in the test datasets.
Here, \textsf{byt5} improves the character error rate (CER) by \input{tolatex/byt5_onlyWords_smallmean_cer_imp.dat}\unskip\% and the word error rate (WER) by \input{tolatex/byt5_onlyWords_smallmean_wer_imp.dat}\unskip\%.

While the focus of this work is on correcting the plain text within our corpus, historical ADS articles also contain inline math and citations.  
Here, we simplify the problem by testing the accuracy of the model on \textit{detecting} these elements in the text.
To proceed, we modify the input and output text by marking these environments with characters that do not appear in the plain text corpus. 
For example, we replace each instance of a SGT or post corrected OCR inline math formula with a single character (\$) and determine how often these characters align in the SGT and predicted OCR. 
The ``\textsf{byt5},full,fixed" row in \autoref{tab:baseline} lists the results of this ``fixed" model, trained on 500k ``fixed" sentences (10k in the validation and test sets).  
Here, the CER and WER improvements have both increased to their highest rates of \input{tolatex/byt5_full_large_fixmean_cer_imp.dat}\unskip\% and \input{tolatex/byt5_full_large_fixmean_wer_imp.dat}\unskip\%, respectively.

To test the model's accuracy on pre-digital OCR, we apply the ``\textsf{byt5},full,fixed" model to  \rt{\input{tolatex/n_historical.dat}\unskip} hand-annotated sentences from the main text of articles in the historical ADS corpus \cite{adsass2012,adsass2015,naiman2022figure,naiman2023digitization}.
When applied to this dataset, the mean improvement, $\langle$I$\rangle$, in CER and WER from correction with the fixed-\textsf{byt5} model (i.e.\ ``\textsf{byt5},full,fixed" for the \arxiv data) are \input{tolatex/byt5_full_large_fix_historicalmean_cer_imp.dat}\unskip\% and \input{tolatex/byt5_full_large_fix_historicalmean_wer_imp.dat}\unskip\%, respectively, as shown in the ``historical,full,fixed" row of \autoref{tab:baseline}. 
While the improvements in CER and WER are more modest than the improvement in the \arxiv dataset, they are nonetheless significantly larger than those from a generic post correction model \cite{huggingfaceyelp} ($\langle $I$ \rangle_{\rm CER}$=\input{tolatex/historical_untrained_mean_cer_imp.dat}\unskip\%, $\langle $I$ \rangle_{\rm WER}$=\input{tolatex/historical_untrained_mean_wer_imp.dat}\unskip\%) or from when \textsf{byt5} is trained on the words from the historical dataset alone ($\langle $I$ \rangle_{\rm CER}$=\input{tolatex/historical_onlyHist_mean_cer_imp.dat}\unskip\%, $\langle $I$ \rangle_{\rm WER}$=\input{tolatex/historical_onlyHist_mean_wer_imp.dat}\unskip\%), both of which result in a large \textit{negative} improvement.
\vspace{-3mm}

\begin{table}
\begin{center}
\begin{tabular}{|c?cccc?cccc|}
    \hline
      & \multicolumn{4}{c?}{CER in \%} & \multicolumn{4}{c|}{WER in \%} \\
     Model  & $\langle$B$\rangle$ & $\langle$A$\rangle$ & $\langle$I$\rangle$ & \% Improved & $\langle$B$\rangle$ & $\langle$A$\rangle$ & $\langle$I$\rangle$ & \% Improved \\
    
    \hline
    \textsf{byt5},words & \input{tolatex/byt5_onlyWords_smallmean_cer_before.dat}\unskip & \input{tolatex/byt5_onlyWords_smallmean_cer_after.dat}\unskip & \input{tolatex/byt5_onlyWords_smallmean_cer_imp.dat}\unskip & \input{tolatex/byt5_onlyWords_smallcerImp.dat}\unskip & \input{tolatex/byt5_onlyWords_smallmean_wer_before.dat}\unskip & \input{tolatex/byt5_onlyWords_smallmean_wer_after.dat}\unskip & \input{tolatex/byt5_onlyWords_smallmean_wer_imp.dat}\unskip & \input{tolatex/byt5_onlyWords_smallwerImp.dat}\unskip\\
    \thickhline 

    \textsf{byt5},full,fixed & \input{tolatex/byt5_full_large_fixmean_cer_before.dat}\unskip & \input{tolatex/byt5_full_large_fixmean_cer_after.dat}\unskip & \input{tolatex/byt5_full_large_fixmean_cer_imp.dat}\unskip & \input{tolatex/byt5_full_large_fixcerImp.dat}\unskip & \input{tolatex/byt5_full_large_fixmean_wer_before.dat}\unskip & \input{tolatex/byt5_full_large_fixmean_wer_after.dat}\unskip & \input{tolatex/byt5_full_large_fixmean_wer_imp.dat}\unskip & \input{tolatex/byt5_full_large_fixwerImp.dat}\unskip \\
    
    \thickhline
    \rt{historical,full,fixed} &  \input{tolatex/byt5_full_large_fix_historicalmean_cer_before.dat}\unskip & \input{tolatex/byt5_full_large_fix_historicalmean_cer_after.dat}\unskip & \input{tolatex/byt5_full_large_fix_historicalmean_cer_imp.dat}\unskip & \input{tolatex/byt5_full_large_fix_historicalcerImp.dat}\unskip & \input{tolatex/byt5_full_large_fix_historicalmean_wer_before.dat}\unskip & \input{tolatex/byt5_full_large_fix_historicalmean_wer_after.dat}\unskip & \input{tolatex/byt5_full_large_fix_historicalmean_wer_imp.dat}\unskip & \input{tolatex/byt5_full_large_fix_historicalwerImp.dat}\unskip \\

    \hline
\end{tabular}
\end{center}
\vspace{-4mm} 
\caption{Mean CER and WER in percent for original datasets, $\langle$B$\rangle$, after post correction with listed models, $\langle$A$\rangle$, and the improvement percent, $\langle$I$\rangle$.  Also shown are the percent of test instances with improvement ($\langle$A$\rangle$$<$$\langle$B$\rangle$) as ``\% Improved". All calculations use the \arxiv dataset except for the last row which uses the historical dataset.}
\label{tab:baseline}
\end{table}

\vspace{-12mm}
\section{Current Limitations \& Future Work}
\vspace{-3mm}

While the full dataset cannot be shared directly (\arxiv administrators, Private communication), we share a subset of our aligned sentences along with analysis notebooks in  GitHub\footnote{\label{github}\url{https://github.com/ReadingTimeMachine/ocr_post_correction}}.  
We are currently working with the \arxiv to make a larger portion of the dataset available to the public.

LaTeX source from $\sim$1990-2010 is known to be difficult to compile due to updates in TeX compilation software \cite{arxivhire} which, in part, lead to the drop of the initial $\sim$65k astronomy articles to $\sim$7k.  Partnership with the \arxiv to support more documents, along with adding support for a wider range of documents (e.g., those with the \textbackslash input command) will increase the dataset size.

While the accuracy of the ``\textsf{byt5},full,fixed" model applied to the historical dataset (``historical,full,fixed") is lower overall, because there is no associated TeX with these historical documents, some ambiguity in the ``ground truth"  is expected (e.g., the phrase ``$\le$90\%" can be written as \verb|$\le$90\%|, \verb|$\le 90$\%| or \verb|$\le 90 \%$| and the meaning of the phrase is unchanged).
Post correction with consideration for these nuances is relegated to future work.

Finally, a larger historical dataset would undeniably enhance our post correction accuracy.  A discussion of the methods used to generate a larger manual dataset is relegated to future work.

\vspace{-5mm}
\subsubsection*{Acknowledgments}
This work is supported by a NASA Astrophysics Data Analysis Program Grant (20-ADAP20-0225). 

\clearpage

\bibliographystyle{splncs04}
\bibliography{references}

\begin{thebibliography}{10}
\providecommand{\url}[1]{\texttt{#1}}
\providecommand{\urlprefix}{URL }
\providecommand{\doi}[1]{https://doi.org/#1}

\bibitem{arxivbulk}
\arxiv bulk downloads. \url{https://info.arxiv.org/help/bulk_data_s3.html},
  accessed: 2022-03-05

\bibitem{arxivhire}
\arxiv hiring and needs. \url{https://info.arxiv.org/hiring/}, accessed:
  2023-07-17

\bibitem{huggingfacebyt5}
Huggingface byt5-small. \url{https://huggingface.co/google/byt5-small},
  accessed: 2023-03-25

\bibitem{huggingfaceyelp}
Huggingface yelpfeast/byt5-base-english-ocr-correction.
  \url{https://huggingface.co/yelpfeast/byt5-base-english-ocr-correction},
  accessed: 2023-07-20

\bibitem{Levenshtein2}
The levenshtein package. \url{https://github.com/maxbachmann/Levenshtein},
  accessed: 2023-05-29

\bibitem{opendetex}
Opendetex. \url{https://github.com/pkubowicz/opendetex}, accessed: 2023-05-29

\bibitem{spacysentence}
The spacy sentence tokenizer. \url{https://spacy.io/api/sentencizer}, accessed:
  2023-05-29

\bibitem{texsoup}
Texsoup. \url{https://github.com/alvinwan/TexSoup}, accessed: 2022-10-30

\bibitem{tikzmark}
The tikzmark package. \url{https://texdoc.org/serve/tikzmark/0}, accessed:
  2023-05-29

\bibitem{adsass2015}
{Accomazzi}, A., {Kurtz}, M.J., {Henneken}, E.A., {Grant}, C.S., {Thompson},
  D., {Chyla}, R., {Holachek}, A., {Sudilovsky}, V., {Murray}, S.S.: {Improved
  Functionality and Curation Support in the ADS}. In: American Astronomical
  Society Meeting Abstracts \#225. American Astronomical Society Meeting
  Abstracts, vol.~225, p. 336.55 (Jan 2015)

\bibitem{ahuja-etal-2022-parsing}
Ahuja, A., Devera, A., Fox, E.A.: Parsing electronic theses and dissertations
  using object detection. In: Proceedings of the first Workshop on Information
  Extraction from Scientific Publications. pp. 121--130. Association for
  Computational Linguistics, Online (Nov 2022),
  \url{https://aclanthology.org/2022.wiesp-1.14}

\bibitem{boros_assessing_2022}
Boros, E., Nguyen, N.K., Lejeune, G., Doucet, A.: Assessing the impact of {OCR}
  noise on multilingual event detection over digitised documents. International
  Journal on Digital Libraries  \textbf{23}(3),  241--266 (Sep 2022).
  \doi{10.1007/s00799-022-00325-2}

\bibitem{icdar2017postcorr}
Chiron, G., Doucet, A., Coustaty, M., Moreux, J.P.: Icdar2017 competition on
  post-ocr text correction. In: 2017 14th IAPR International Conference on
  Document Analysis and Recognition (ICDAR). vol.~01, pp. 1423--1428 (2017).
  \doi{10.1109/ICDAR.2017.232}

\bibitem{eichhorn_new_2002}
Eichhorn, G., Accomazzi, A., Grant, C.S., Kurtz, M.J., Rey~Bacaicoa, V.,
  Murray, S.S.: New {Data} and {Search} {Features} in the {NASA} {ADS}
  {Abstract} {Service} p.~1298 (Mar 2002),
  \url{https://ui.adsabs.harvard.edu/abs/2002LPI....33.1298E}, conference Name:
  Lunar and Planetary Science Conference ADS Bibcode: 2002LPI....33.1298E

\bibitem{etter_synthetic_2019}
Etter, D., Rawls, S., Carpenter, C., Sell, G.: A {Synthetic} {Recipe} for
  {OCR}. In: 2019 {International} {Conference} on {Document} {Analysis} and
  {Recognition} ({ICDAR}). pp. 864--869. IEEE, Sydney, Australia (Sep 2019).
  \doi{10.1109/ICDAR.2019.00143}

\bibitem{overproofdata}
Evershed, J., Fitch, K.: Correcting noisy ocr: Context beats confusion. In:
  Proceedings of the First International Conference on Digital Access to
  Textual Cultural Heritage. p. 45–51. DATeCH '14, Association for Computing
  Machinery, New York, NY, USA (2014). \doi{10.1145/2595188.2595200}

\bibitem{ginev_latexml_2014}
Ginev, D., Miller, B.R.: Latexml 2012-a year of latexml. In: Intelligent
  Computer Mathematics: MKM, Calculemus, DML, and Systems and Projects 2013,
  Held as Part of CICM 2013, Bath, UK, July 8-12, 2013. Proceedings 6. pp.
  335--338. Springer (2013)

\bibitem{spacy2}
Honnibal, M., Montani, I.: {spaCy 2}: Natural language understanding with
  {B}loom embeddings, convolutional neural networks and incremental parsing
  (2017), to appear

\bibitem{jiang2021gutenberg}
Jiang, M., Hu, Y., Worthey, G., Dubnicek, R.C., Capitanu, B., Kudeki, D.,
  Downie, J.S., et~al.: The gutenberg-hathitrust parallel corpus: A real-world
  dataset for noise investigation in uncorrected ocr texts. In: iConference
  2021 (Poster) (2021)

\bibitem{scanbankthesis}
Kahu, S.Y.: Figure Extraction from Scanned Electronic Theses and Dissertations.
  Master's thesis, Virginia Tech (2020)

\bibitem{krishnan2016generating}
Krishnan, P., Jawahar, C.: Generating synthetic data for text recognition.
  arXiv preprint arXiv:1608.04224  (2016)

\bibitem{le2017using}
Le, T.A., Baydin, A.G., Zinkov, R., Wood, F.: Using synthetic data to train
  neural networks is model-based reasoning. In: 2017 international joint
  conference on neural networks (IJCNN). pp. 3514--3521. IEEE (2017)

\bibitem{levenshtein}
{Levenshtein}, V.I.: {Binary Codes Capable of Correcting Deletions, Insertions
  and Reversals}. Soviet Physics Doklady  \textbf{10}, ~707 (Feb 1966)

\bibitem{li_tablebank_2020}
Li, M., Cui, L., Huang, S., Wei, F., Zhou, M., Li, Z.: {TableBank}: {A}
  {Benchmark} {Dataset} for {Table} {Detection} and {Recognition} (Jul 2020),
  \url{http://arxiv.org/abs/1903.01949}, arXiv:1903.01949 [cs]

\bibitem{maskrcnnDocbank}
Li, M., Xu, Y., Cui, L., Huang, S., Wei, F., Li, Z., Zhou, M.: Docbank: A
  benchmark dataset for document layout analysis. In: Proceedings of the 28th
  International Conference on Computational Linguistics. pp. 949--960 (2020)

\bibitem{liu2020mBART}
Liu, Y., Gu, J., Goyal, N., Li, X., Edunov, S., Ghazvininejad, M., Lewis, M.,
  Zettlemoyer, L.: {Multilingual Denoising Pre-training for Neural Machine
  Translation}. Transactions of the Association for Computational Linguistics
  \textbf{8},  726--742 (11 2020). \doi{10.1162/tacl\_a\_00343}

\bibitem{maheshwari_benchmark_2022}
Maheshwari, A., Singh, N., Krishna, A., Ramakrishnan, G.: A {Benchmark} and
  {Dataset} for {Post}-{OCR} text correction in {Sanskrit} (Nov 2022).
  \doi{10.48550/arXiv.2211.07980}, arXiv:2211.07980 [cs]

\bibitem{mayernik2017}
Mayernik, M.S., Hart, D.L., Maull, K.E., Weber, N.M.: Assessing and tracing the
  outcomes and impact of research infrastructures. Journal of the Association
  for Information Science and Technology  \textbf{68}(6),  1341--1359 (2017).
  \doi{https://doi.org/10.1002/asi.23721}

\bibitem{naiman2023digitization}
Naiman, J.P., Williams, P.K., Goodman, A.: The digitization of historical
  astrophysical literature with highly localized figures and figure captions.
  International Journal on Digital Libraries pp. 1--21 (2023)

\bibitem{naiman2022figure}
Naiman, J., Williams, P.K., Goodman, A.: Figure and figure caption extraction
  for mixed raster and vector pdfs: Digitization of astronomical literature
  with ocr features. In: Linking Theory and Practice of Digital Libraries: 26th
  International Conference on Theory and Practice of Digital Libraries, TPDL
  2022, Padua, Italy, September 20--23, 2022, Proceedings. pp. 52--67. Springer
  (2022)

\bibitem{nguyen_deep_2019}
Nguyen, T.T.H., Jatowt, A., Coustaty, M., Nguyen, N.V., Doucet, A.: Deep
  {Statistical} {Analysis} of {OCR} {Errors} for {Effective} {Post}-{OCR}
  {Processing}. In: 2019 {ACM}/{IEEE} {Joint} {Conference} on {Digital}
  {Libraries} ({JCDL}). pp. 29--38 (Jun 2019). \doi{10.1109/JCDL.2019.00015}

\bibitem{adsass2012}
{Pepe}, A., {Goodman}, A., {Muench}, A.: {The ADS All-Sky Survey}. In:
  {Ballester}, P., {Egret}, D., {Lorente}, N.P.F. (eds.) Astronomical Data
  Analysis Software and Systems XXI. Astronomical Society of the Pacific
  Conference Series, vol.~461, p.~275 (Sep 2012)

\bibitem{pfahler_self-supervised_2022}
Pfahler, L., Morik, K.: Self-{Supervised} {Pretraining} of {Graph} {Neural}
  {Network} for the {Retrieval} of {Related} {Mathematical} {Expressions} in
  {Scientific} {Articles} (Aug 2022), \url{http://arxiv.org/abs/2209.00446},
  arXiv:2209.00446 [cs]

\bibitem{ramirez2022post}
Ramirez-Orta, J.A., Xamena, E., Maguitman, A., Milios, E., Soto, A.J.: Post-ocr
  document correction with large ensembles of character sequence-to-sequence
  models. In: Proceedings of the AAAI Conference on Artificial Intelligence.
  vol.~36, pp. 11192--11199 (2022)

\bibitem{rensynthOCRdata2016}
{Ren}, X., {Chen}, K., {Sun}, J.: {A CNN Based Scene Chinese Text Recognition
  Algorithm With Synthetic Data Engine}. arXiv e-prints arXiv:1604.01891 (Apr
  2016). \doi{10.48550/arXiv.1604.01891}

\bibitem{rigaud_icdar_nodate}
Rigaud, C., Doucet, A., Coustaty, M., Moreux, J.P.: {ICDAR} 2019 {Competition}
  on {Post}-{OCR} {Text} {Correction}

\bibitem{icdar2019postcorr}
Rigaud, C., Doucet, A., Coustaty, M., Moreux, J.P.: Icdar 2019 competition on
  post-ocr text correction. In: 2019 International Conference on Document
  Analysis and Recognition (ICDAR). pp. 1588--1593 (2019).
  \doi{10.1109/ICDAR.2019.00255}

\bibitem{saier2019bibliometric}
Saier, T., F{\"a}rber, M.: Bibliometric-enhanced arxiv: A data set for
  paper-based and citation-based tasks. In: BIR@ ECIR. pp. 14--26 (2019)

\bibitem{unarxiv2022}
{Saier}, T., {Krause}, J., {F{\"a}rber}, M.: {unarXive 2022: All arXiv
  Publications Pre-Processed for NLP, Including Structured Full-Text and
  Citation Network}. arXiv e-prints arXiv:2303.14957 (Mar 2023).
  \doi{10.48550/arXiv.2303.14957}

\bibitem{sandy2017}
Sandy, H.M., Mitchell, E., Corrado, E.M., Budd, J., West, J.D., Bossaller, J.,
  VanScoy, A.: Making a case for open research: Implications for
  reproducibility and transparency. Proceedings of the Association for
  Information Science and Technology  \textbf{54}(1),  583--586 (2017).
  \doi{https://doi.org/10.1002/pra2.2017.14505401079}

\bibitem{schmitt-koopmann_accessible_2022}
Schmitt-Koopmann, F.M., Huang, E.M., Darvishy, A.: Accessible {PDFs}:
  {Applying} {Artificial} {Intelligence} for {Automated} {Remediation} of
  {STEM} {PDFs}. In: Proceedings of the 24th {International} {ACM} {SIGACCESS}
  {Conference} on {Computers} and {Accessibility}. pp.~1--6. {ASSETS} '22,
  Association for Computing Machinery, New York, NY, USA (Oct 2022).
  \doi{10.1145/3517428.3550407}

\bibitem{astrobigdata2}
{Smith}, L., {Arcand}, K., {Smith}, R., {Bookbinder}, J., {Smith}, J.:
  {Capturing the many faces of an exploded star: communicating complex and
  evolving astronomical data}. JCOM Journal of Science Communication
  \textbf{16}, 16050202 (Nov 2017). \doi{10.22323/2.16050202}

\bibitem{tesseract}
Smith, R.: An overview of the tesseract ocr engine. In: Proceedings of the
  Ninth International Conference on Document Analysis and Recognition - Volume
  02. p. 629–633. ICDAR '07, IEEE Computer Society, USA (2007)

\bibitem{sohmen2018figures}
Sohmen, L., Charbonnier, J., Bl{\"u}mel, I., Wartena, C., Heller, L.: Figures
  in scientific open access publications. In: International Conference on
  Theory and Practice of Digital Libraries. pp. 220--226. Springer (2018)

\bibitem{springmann2018ground}
Springmann, U., Reul, C., Dipper, S., Baiter, J.: Ground truth for training ocr
  engines on historical documents in german fraktur and early modern latin.
  Journal for Language Technology and Computational Linguistics
  \textbf{33}(1),  97--114 (2018)

\bibitem{astrobigdata1}
Stephens, Z.D., Lee, S.Y., Faghri, F., Campbell, R.H., Zhai, C., Efron, M.J.,
  Iyer, R., Schatz, M.C., Sinha, S., Robinson, G.E.: Big data: Astronomical or
  genomical? PLOS Biology  \textbf{13}(7),  1--11 (07 2015).
  \doi{10.1371/journal.pbio.1002195}

\bibitem{van2020assessing}
Strien, D., Beelen, K., Coll~Ardanuy, M., Hosseini, K., Mcgillivray, B.,
  Colavizza, G.: Assessing the impact of ocr quality on downstream nlp tasks.
  SCITEPRESS-Science and Technology Publications (02 2020).
  \doi{10.5220/0009169004840496}

\bibitem{tafti2016ocr}
Tafti, A.P., Baghaie, A., Assefi, M., Arabnia, H.R., Yu, Z., Peissig, P.: Ocr
  as a service: an experimental evaluation of google docs ocr, tesseract, abbyy
  finereader, and transym. In: Advances in Visual Computing: 12th International
  Symposium, ISVC 2016, Las Vegas, NV, USA, December 12-14, 2016, Proceedings,
  Part I 12. pp. 735--746. Springer (2016)

\bibitem{urban1986introduction}
Urban, M.: An introduction to LATEX. TEX users group (1986)

\bibitem{xue_byt5_2022}
Xue, L., Barua, A., Constant, N., Al-Rfou, R., Narang, S., Kale, M., Roberts,
  A., Raffel, C.: {ByT5}: {Towards} a {Token}-{Free} {Future} with
  {Pre}-trained {Byte}-to-{Byte} {Models}. Transactions of the Association for
  Computational Linguistics  \textbf{10},  291--306 (2022).
  \doi{10.1162/tacl\_a\_00461}, place: Cambridge, MA Publisher: MIT Press

\bibitem{zaytsev2015hathitrust}
Zaytsev, A.: Hathitrust and a mission for accessibility. Journal of Electronic
  Publishing  \textbf{18}(3) (2015)

\bibitem{zharikov_ddi-100-2020}
Zharikov, I., Nikitin, F., Vasiliev, I., Dokholyan, V.: {DDI}-100: {Dataset}
  for {Text} {Detection} and {Recognition}. In: Proceedings of the 2020 4th
  {International} {Symposium} on {Computer} {Science} and {Intelligent}
  {Control}. pp.~1--5 (Nov 2020). \doi{10.1145/3440084.3441192},
  arXiv:1912.11658 [cs]

\bibitem{zhu_novel_2016}
Zhu, W., Liu, Y., Hao, L.: A {Novel} {OCR} {Approach} {Based} on {Document}
  {Layout} {Analysis} and {Text} {Block} {Classification}. In: 2016 12th
  {International} {Conference} on {Computational} {Intelligence} and {Security}
  ({CIS}). pp. 91--94 (Dec 2016). \doi{10.1109/CIS.2016.0029}

\end{thebibliography}

\end{document}